\documentclass[11pt,twoside]{article}

\usepackage{asp2014}

\aspSuppressVolSlug
\resetcounters

\begin{document}

\title{A distributed computing infrastructure for LOFAR Italian community }

\author{Giuliano~Taffoni,$^1$ Ugo~Becciani,$^2$ Annalisa~Bonafede,$^{3,4}$ Etienne~Bonnassieux,$^{^4}$  Gianfranco Brunetti,$^3$  Marisa~Brienza,$^{3,4}$ Claudio~Gheller,$^3$ Stefano~A.~Russo,$^1$ and Fabio~Vitello$^3$
\affil{$^1$INAF -- Astronomical Observatory of  Trieste, Trieste,  Italy; \email{giuliano.taffoni@inaf.it}}
\affil{$^2$INAF -- Astrophysical Observatory of Catania, Catania, Italy}
\affil{$^3$INAF -- Institute for Radio Astronomy, Bologna, Italy}
\affil{$^4$DIFA-- Universit\'a di Bologna, Bologna, Italy}
}

\paperauthor{Giuliano~Taffoni}{giuliano.taffoni@inaf.it}{0000-0002-4211-6816}{INAF}{Astronomical Observatory of Trieste}{Trieste}{}{34131}{Italy}
\paperauthor{Ugo~Becciani}{ugo.becciani@inaf.it}{https://orcid.org/0000-0002-4389-8688}{INAF- OACt}{Osservatorio Astrofisico di Catania}{Catania}{}{95123}{Italy}
\paperauthor{Annalisa~Bonafede}{annalisa.bonafede@unibo.it}{0000-0002-5068-4581}{Universit\'a di Bologna}{Dipartimento di Fisica e Astronomia}{Bologna}{BO}{40131}{Italy}
\paperauthor{Etienne~Bonnassieux}{etienne.bonnassieux@unibo.it}{0000-0003-2312-3508}{Universit\'a di Bologna}{Dipartimento di Fisica e Astronomia}{Bologna}{BO}{40131}{Italy}
\paperauthor{Gianfranco~Brunetti}{gianfranco.brunetti@inaf.it}{0000-0003-4195-8613}{INAF}{Institute of Radioastronomy}{}{Bologna}{40129}{Italy}
\paperauthor{Marisa~Brienza}{m.brienza@ira.inaf.it}{0000-0003-4120-9970}{Universit\'a di Bologna}{Dipartimento di Fisica e Astronomia}{Bologna}{BO}{40131}{Italy}
\paperauthor{Claudio~Gheller}{claudio.gheller@inaf.it}{0000-0003-1063-3541}{INAF}{Institute of Radioastronomy}{}{Bologna}{40129}{Italy}
\paperauthor{Stefano~Alberto~Russo}{stefano.russo@inaf.it}{0000-0003-4487-6752}{INAF}{Astronomical Observatory of Trieste}{Trieste}{}{34131}{Italy}
\paperauthor{Fabio~Vitello}{fabio.vitello@inaf.it}{0000-0003-2203-3797}{INAF}{Institute of Radioastronomy}{}{Bologna}{40129}{Italy}


\begin{abstract}
The LOw Frequency ARray is a low-frequency radio interferometer composed by observational stations spread across Europe and it is the largest precursor of SKA in terms of effective area and generated data rates.
In 2018, the Italian community officially joined LOFAR project, and it deployed a distributed computing and storage infrastructure dedicated to LOFAR data analysis. The infrastructure is based on 4 nodes distributed in different Italian locations and it offers services  for pipelines execution, storage of final and intermediate results and  support for the use of the software and infrastructure.  As the analysis of the LOw Frequency ARray data requires a very complex computational procedure, a container--based approach has been adopted to distribute software environments to the different computing resources. A science platform approach is used to facilitate interactive access to computational resources. In this poster we describe the architecture and main features of the infrastructure. 
\end{abstract}



\section{Introduction}
LOFAR (Low--Frequency Array) \citep{2013A&A...556A...2V} is a phased array radio telescope consisting of a large number of stations located in various countries in Europe\footnote{38 Dutch stations plus 14 stations spread in 8 different countries and a new one under development in Italy}. It operates between 10 and 240 MHz allowing detailed sensitive high--resolution studies of the low--frequency radio sky in various astrophysical fields. With its sensitivity LOFAR is one of then primary SKA low frequency precursor.

In 2018, INAF became member of the International LOFAR Consortium. 
INAF promoted the setting up of the LOFAR--IT consortium of which it is a member and the chair institution. 
The consortium involves also the Department of Physics of the University of Turin and aims at providing to the Italian scientists a framework to facilitate their participation to the LOFAR International Collaboration.

The final goal of this effort  is to optimise the scientific impact of the Italian community, through supporting the  involvement in the LOFAR Key Projects, the submission of proposals to the periodic calls and the access to data, analysis software and platforms.

\section{LOFAR--IT Data Working Group}

To achieve its objectives, the LOFAR--IT Consortium promoted the development of a National Distributed Computing Infrastructure  managed by a DATA Working Group (DWG).

DWG is aiming at: (i) designing, implementing and coordinating the Italian hardware and software  infrastructure for data calibration,  reduction and analysis; (ii) coordinating the installation, configuration and management of specific software and pipelines for the reduction and analysis of LOFAR data; (iii) providing technical support to LOFAR--IT users through testing, verification, optimization and development of the different pipelines; (iv) collaborating with the LOFAR International Telescope for the development of the telescope and archiving software; (v) codes and pipelines optimization and re-design, including parallelization of algorithms and software, to address new computing platforms.

The DWG involves a  group of researchers and post-docs  that collaborate to carry oout efficiently all the tasks previously mentioned. It is organized into 3  tasks: (Task1) User support and training,   (Task 2) Infrastructure, (Task 3) Pipelines and algorithms.
Tasks are not independent and  cooperate to maintain the infrastructure, support the community and provides software and pipelines.

\section{LOFAR--IT Infrastructure}

The LOFAR--IT infrastructure is a distributed computing environment  based on 4 computing centres located at INAF--OATs and
INAF--OACt, INAF--IRA and University of Turin.

\smallskip
\noindent \textbf{The HOTCAT Cluster at INAF--OATs} \citep{2020ASPC..527..303B} is an HPC infrastructure 
with 1400 INTEL cores.  It includes two different types of computing
nodes: Light (INTEL E5-4627v3 256GB), Fat (INTEL Xeon(R) Gold 5118 512GB). The system has two types of storage components:80TB of archive storage (an expandable general--purpose storage), 500TB of scratch storage (a high performance appliance for high-bandwidth data access based on BeeGFS).  
The storage and computing  interconnect is Infiniband ConnectX  Pro Dual QSFP+ 54Gbs 
(Figures \ref{fig1}), 
the management network is 1Gbs Ethernet. The cluster is based on SLURM resource manager.  

\smallskip
\noindent  \textbf{INAF--OACt cluster} is a HTC infrastructure with 384 INTEL cores\citep{2020ASPC..527..307T}. The nodes are INTEL E5-2620V2 with 5.2GB RAM per Core (1TB RAM total) and 70 TB distributed storage (NFS). The
interconnect is a 10 Gbs Ethernet network. The cluster resource manager is  PBS-pro.
\articlefigure[width=.7\textwidth]{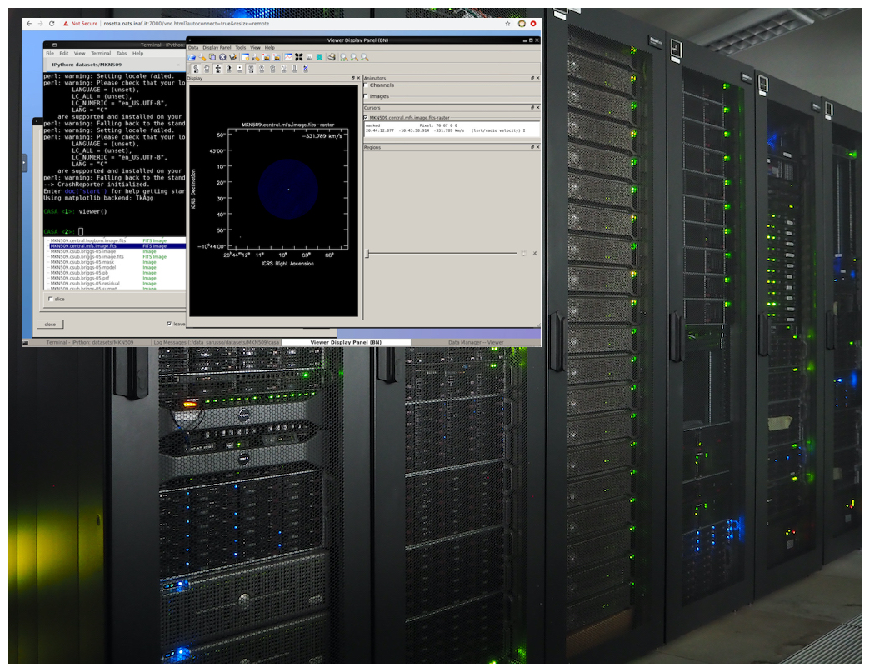}{fig1}{INAF-OATs server room and HOTCAT cluster and a ROSETTA science platform screenshot running a CASA analysis.}

\smallskip
\noindent  \textbf{INAF--IRA infrastructure} is a cluster consisting of Intel Xeon Gold 6130 nodes with 384GB (light) and 512GB (Fat) of RAM, each node is equipped with a local filesystem of 10TB used as local scratch and Lustre parallel filesystem of 240TB. The interconnect is a 10 Gbs Ethernet network.

\smallskip
\noindent  \textbf{The Open Computing Cluster for Advanced data Manipulation} (OCCAM) is a multipurpose flexible HPC/HTC cluster \citep{17:occam:chep}. It includes three different types of computing
nodes: Light (INTEL E5-2680v3 128GB), Fat (INTEL E7-4830v3 768GB) and GPU (light nodes with 2 nVidia K40).  The infrastructure has two storage appliances (750TB Archive and 250TB Scratch). All components,
both for computing and storage, are connected by three different network links including InfiniBand 56 Gb/s FDR. OCCAM is a container based  infrastructure where Docker containers are used to run  user-defined images in the system and  to. partition the nodes.

A consistent fraction of those resources  (between 10 and 20 per cent according to the site) is dedicated to LOFAR community.

\section{Software and services}
The LOFAR--IT distributed infrastructure is aimed at providing a flexible, configurable and extendable infrastructure to cater to a wide range of different LOFAR  pipelines and use cases. Furthermore, it serves as a platform for  software, pipeline and algorithms development.
It is  equipped with more than 60 software environment for Astronomical data reduction and analysis, and  tools for software
development, profiling and debugging. 

A distributed heterogeneous system like this poses a number of challenges related to the resource assignment and   
availability, which in turn affect methods and means to allocate, manage, optimize, and  monitor the HW resources.
This is complicated by the fact that LOFAR data reduction and analysis pipelines requires extremely long execution times ($~150$ continuous processing hours) and 
do not implement check--pointing. Furthermore, they produce large amount of data in the various stages of the data 
reduction (e.g. between 5 and 15 TB for an interferometric dataset of 8h observing time) to archive till the end of the scientific analysis.

We have defined an operational workflow that identifies for each user, since the proposal submission, the requirements in terms of infrastructure occupancy and support requested. Users are assigned to one of the sites and supported by a group of experts that guide them in the use of the computing resources and software until the scientific analysis is completed.  To provide the same software and software environment  for the analysis independently of the platform, we use only containerized environments (Docker and Singularity).  The use of containers not only simplifies software distribution and upgrades, but is also guarantee the reproducibility of the scientific analysis.

Open Container Initiative compliant images are created with Docker and stored in a private image registry\footnote{https://www.ict.inaf.it/gitlab/lofarit}. Images of public software are freely available; Key Projects private software access is protected. 
As the data analysis requires some level of interactivity, a science platform has been developed to allow interactive   computing on our clusters (https://rosetta.oats.inaf.it/).

\section{Conclusion} 
We present the Italian Computing Infrastructure for LOFAT--IT. This is a distributed environment based on containers and 
a science platform. Policies have been defined for using the resources to optimize the support and the use of the infrastructure.

More than $100$ referred papers has been published by the Italian community using Key Projects data since the setting up of the LOFAR--IT consortium. 

The recent acquisition of a new set of resources, will dramatically improve the infrastructure in terms of computing resource: $~2700$ INTEL Broadwell cores will be installed at INAF--IRA fully dedicated to LOFAR--IT.
Those resources will be made available also to the LOFAT International Community to support Key Projects.

\acknowledgements ESCAPE -- the European Science Cluster of Astronomy \& Particle Physics ESFRI Research Infrastructures --
has received funding from the European Union's Horizon 2020 research and innovation programme under the
Grant Agreement n. 824064. Marisa~Brienza acknowledges support from the ERC-Stg ``DRANOEL", no. 714245, from the ERC-Stg ``MAGCOW", no. 714196.

\bibliography{X7-012}  


\end{document}